\documentclass[letterpaper, 10 pt, conference]{ieeeconf}   
\IEEEoverridecommandlockouts                               
\usepackage{optidef}

\usepackage{amssymb,amsmath,color,subfigure}
\usepackage{graphicx}
\usepackage{xspace,stackrel,hyperref}
\usepackage{wrapfig}
\usepackage{mathtools}
\usepackage{tikz,pgfplots}
\usepackage{etoolbox}
\usepackage{autonum}
\usepackage{dsfont}
\usepackage{etoolbox}

\usepackage{multirow}
\usepackage{cite}

\usepgfplotslibrary{fillbetween}
\usepgfplotslibrary{groupplots}

\usepackage{lipsum,adjustbox}

\definecolor{mycolor4}{RGB}{230,97,1}
\definecolor{mycolor2}{RGB}{178,171,210}
\definecolor{mycolor3}{RGB}{253,184,99}
\definecolor{mycolor1}{RGB}{94,60,153}

\makeatletter 
\pretocmd\@bibitem{\color{black}\csname keycolor#1\endcsname}{}{\fail}
\newcommand\citecolor[1]{\@namedef{keycolor#1}{\color{blue}}}
\makeatother

\DeclareMathAlphabet{\mathcal}{OMS}{cmsy}{m}{n}

\def\beq{\begin{equation}}
\def\eeq{\end{equation}}

\newcommand{\mc}{\mathcal}

\newcommand{\R}{\mathds{R}}

\newcommand{\defineas}{\coloneqq}

\definecolor{mycolor1}{RGB}{230,97,1}
\definecolor{mycolor2}{RGB}{178,171,210}
\definecolor{mycolor3}{RGB}{253,184,99}
\definecolor{mycolor4}{RGB}{94,60,153}
\definecolor{mycolor5}{rgb}{0,0,0}

\tikzset{
  pics/car/.style args={#1}{
     code={
     \begin{scope}[scale=0.15]
      \shade[top color=#1, bottom color=white, shading angle={135}]
        [draw=black,fill=red!20,rounded corners=0.2ex] (1.5,.5) -- ++(0,1) -- ++(1,0.3) --  ++(3,0) -- ++(1,0) -- ++(0,-1.3) -- (1.5,.5) -- cycle;
    \draw[ rounded corners=0.5ex,fill=black!20!blue!20!white]  (2.5,1.8) -- ++(1,0.7) -- ++(1.6,0) -- ++(0.6,-0.7) -- (2.5,1.8);
    \draw[thick]  (4.2,1.8) -- (4.2,2.5);
    \draw[draw=black,fill=gray!50,thick] (2.75,.5) circle (.5);
    \draw[draw=black,fill=gray!50,thick] (5.5,.5) circle (.5);
    \end{scope}
     }
  }
}

\newtheorem{theorem}{Theorem}

\newtheorem{definition}{Definition}
\newtheorem{lemma}{Lemma}
\newtheorem{remark}{Remark}

\newtoggle{full_version}

\toggletrue{full_version}

\title{\LARGE \bf
A Blotto Game Approach to Ride-hailing Markets with Electric Vehicles
}

\author{Marko Maljkovic, Gustav Nilsson, and Nikolas Geroliminis
\thanks{M.~Maljkovic, G.~Nilsson, and N.~Geroliminis are with the School of Architecture, Civil and Environmental Engineering, École Polytechnique Fédérale de Lausanne (EPFL), 1015 Lausanne, Switzerland. {\tt\small \{marko.maljkovic, gustav.nilsson, nikolas.geroliminis\}@epfl.ch}.}%
\thanks{This work was supported by the Swiss National Science Foundation under NCCR Automation, grant agreement 51NF40\_180545.}
\iftoggle{full_version}{}{\thanks{An extended version containing all the proofs is available at \url{http://?????????????}}}%
}

\begin{document}

\maketitle
\thispagestyle{empty}
\pagestyle{empty}

\begin{abstract}
When a centrally operated ride-hailing company considers to enter a market already served by another company, it has to make a strategic decision about how to distribute its fleet among different regions in the area. This decision will be influenced by the market share the company can secure and the costs associated with charging the vehicles in each region, all while competing with the company already operating in the area. In this paper, we propose a Colonel Blotto-like game to model this decision-making. For the class of games that we study, we first prove the existence and uniqueness of a Nash Equilibrium. Subsequently, we provide its general characterization and present an algorithm for computing the ones in the feasible set’s interior. Additionally, for a simplified scenario involving two regions, which would correspond to a city area with a downtown and a suburban region, we also provide a method to check for the equilibria on the feasible set's boundary. Finally, through a numerical case study, we illustrate the impact of charging prices on the position of the Nash equilibrium.
\end{abstract}

\section{Inroduction}

Ride-hailing services have revolutionized urban transportation, offering convenience, cost-efficiency, and flexibility to their ever-increasing number of customers. With the simultaneous need to curb the environmental impact of traditional gas-powered vehicles, the rapid adoption of electric vehicles (EVs)~\cite{iea} in the ride-hailing market has become a necessity as they govern the transition to sustainable and environmentally friendly modes of transportation. However, the integration of EVs into the ride-hailing ecosystem presents various challenges, from managing the charging of the vehicles to optimizing fleet operations~\cite{9318522,EVsCharging,article4}. In particular, when a centrally operated ride-hailing company wants to enter an existing market already served by another company, it has to make a strategic decision on how to distribute vehicles among different regions in order to run a profitable business. This decision will be primarily influenced by the expected ride-hailing market share for the company and the anticipated operational costs, all within a competitive environment shared with other service providers~\cite{9102356,article5,Paccagnan2016a}. The company's objective is to maximize its influence in the regions and effectively meet the heterogeneously distributed ride-hailing demand while factoring in different charging prices offered by existing charging infrastructure in the area.     

\begin {figure}
\centering
\begin{adjustbox}{max height=0.55\textwidth, max width=0.45\textwidth}

    \begin{tikzpicture}[scale=0.8]

    \node[ text=black, shape=rectangle](ch1) at (-8.0, 1.2){Company A};
    \node (pic2) at (-8.0, 0.0) {\includegraphics[width=.09\textwidth]{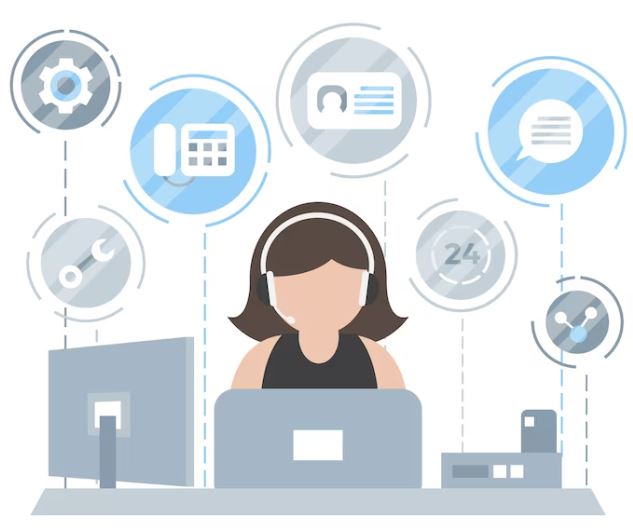}};
    \draw[draw=mycolor1, rounded corners] (-9.5, 1.5) rectangle (-6.5,-1.5);

    \node[text=black, shape=rectangle](ch2) at (8.0, 1.2){Company B};
    \node (pic1) at (8.0, 0.0) {\includegraphics[width=.09\textwidth]{figures/operator.png}};
    \draw[draw=mycolor2, rounded corners] (6.5, 1.5) rectangle (9.5,-1.5);

    \node (myfirstpic) at (0.0, 0) {\includegraphics[width=.4\textwidth]{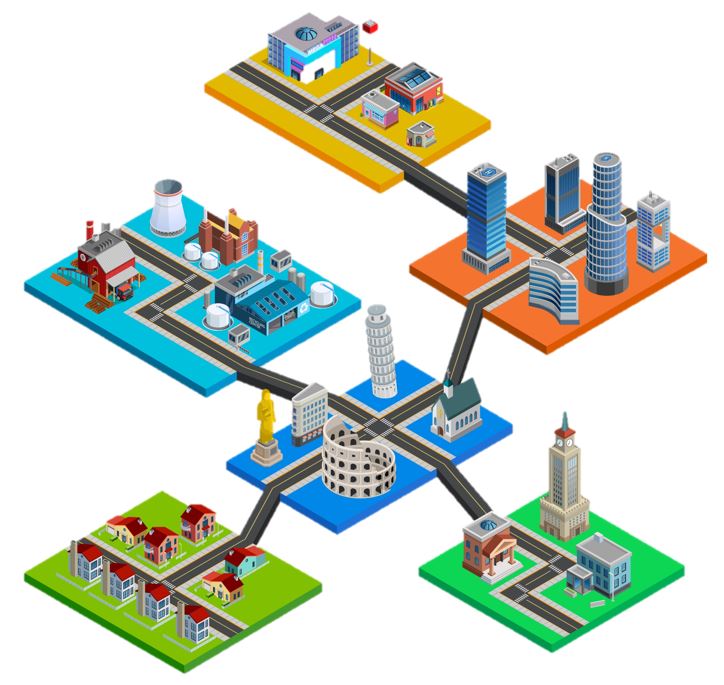}};

    \draw[dashed, draw=black, rounded corners] (-4.5, -4.8) rectangle (4.5,4.8);
    
    \fill[left color =mycolor1, right color=white] (-4.6,-4.8) -- (-4.6,4.8) -- (-6.5,0.0) -- cycle;
    \fill[right color=mycolor2, left color= white] (4.6,-4.8) -- (4.6,4.8) -- (6.5, 0.0) -- cycle;

    \node[] (ba) at (-8.0, -1.2){$\mathbf{X}_a$ vehicles};
    \node[] (bb) at (8.0, -1.2){$\mathbf{X}_b$ vehicles};


    \node[shape=circle] (j1) at (0.8, 3.9){$J_1$};
    \node[shape=circle] (j2) at (-3.4, 1.9){$J_6$};
    \node[shape=circle] (j3) at (3.7, 2.3){$J_2$};
    
    \node[shape=circle] (j4) at (-3.5, -1.8){$J_5$};
    \node[shape=circle] (j5) at (0.7,0.3){$J_3$};
    \node[shape=circle] (j6) at (3.5, -1.8){$J_4$};
            
{\tiny }
\end{tikzpicture}
\end{adjustbox}
    \caption{Representation of the Blotto game setup with two companies and six 'battlefields' $\mc J=\{J_1,J_2,J_3,J_4,J_5,J_6\}$.} \vspace{-2em}
\label{fig:problem}
\end{figure}

Tackling the intricate dynamics of ride-hailing markets with EVs and multiple competing companies has already been explored in existing literature~\cite{Hierarchical, ECC2023, maljkovic2023learning}. However, these works primarily focus on determining optimal charging prices in different regions with the aim to incentivize an exogenously determined societal optimum that could help reduce congestion in demand-attractive regions and total queuing times at the charging stations resulting from their limited capacity. The proposed hierarchical, Stackelberg-like, game structures assume that company operators split their fleets so as to minimize the total operational cost that, among others, includes the cost of charging and the expected profit earned from operating in different regions. However, the expected profit does not model the interactions with other companies. Instead, it has been simplified to a parametrized linear function of the personal decision vector, providing only an estimate based on historical data. 

We propose a novel approach inspired by the principles of Blotto games~\cite{Borel1953TheTO}, commonly perceived as strategic allocation games in which players distribute resources or troops across multiple battlefields~\cite{9782697, 9993253, KIM20181}. In this paper, a Blotto-like model is used to describe how companies split the share of ride-hailing users. Moreover, their decisions will be influenced by the cost of charging in each region, which inherently impacts the overall profitability of the company. Within the realm of the ride-hailing market, the framework depicted in Figure~\ref{fig:problem} conceptualizes each region as a distinct 'battlefield' equipped with charging stations, where the ride-hailing companies can deploy their EVs in a strategic competition to gain market share and secure cheaper charging. To the best of our knowledge, our model is the first one to take into account the cost of charging and to introduce the concept of customer abandonments. In the existing literature, the market share acquisition is typically modeled using classical Blotto games with Tullock contest success functions~\cite{OSORIO2013164, KIM20181, LI2022102771}, meaning that the players share the whole profit that can be earned in a particular region. In our framework, we account for the fact that it is not unlikely that the customers decide to abandon their service requests due to long waiting times caused by insufficient demand coverage. As a result, although the overall profit potential for the company in a specific region increases with the number of deployed vehicles, a portion of it will inevitably be forfeited depending on the companies' participation within the region.   

Apart from the novel model, the main contribution of this paper involves establishing the existence and uniqueness of the Nash equilibrium of the proposed game. For a general setup, we analyze the best-response optimization problems of the companies which results in a method to analytically compute the Nash equilibrium if it is located in the feasible set's interior. Moreover, for a two-region setup, we are also able to provide steps to analytically compute the ones on the boundary. Finally, through a numerical case study, we illustrate the influence of different charging prices on the attained Nash equilibrium (NE) of the proposed model.

The paper is outlined as follows: the rest of this section introduces the basic notation. In Section~\ref{sec:model}, we outline the general problem statement and present the structure of the electric ride-hailing market. In Section~\ref{sec:NE}, we then present our main methodological and theoretical results. Finally, we conclude the paper with Sections~\ref{sec:example} and~\ref{sec:conclusion}, where we illustrate our model in a numerical case study and propose ideas for future research.

\textit{Notation:}  Let $\R$ denote the set of real numbers and $\R_{\geq0}$ the set of non-negative reals. Let $\mathbf{1}_{m}$ denote the vector of all ones of length $m$. For a finite set $\mc A$, we let $\R_{(\geq0)}^{\mc A}$ denote the set of (non-negative) real vectors indexed by the elements of $\mc A$ and $\left|\mc A\right|$ the cardinality of $\mc A$. For finite sets $\mc A$, $\mc B$ and a set of $|\mc B|$ vectors $x^i\in \R_{\geq0}^{\mc A}$, we define $x \defineas \text{col}((x^{i})_{i\in \mc B})\in \R^{|\mc A||\mc B|}$ to be their concatenation. For a vector $x\in\R^n$, we let $\text{diag}(x)\in\R^{n\times n}$ denote a diagonal matrix whose elements on the diagonal correspond to vector $x$.

\section{Problem statement}\label{sec:model}
We consider a scenario in which two players, in this case ride-hailing companies $a$ and $b$, are interested in distributing their fleets of respective sizes $\mathbf{X}_a$ and $\mathbf{X}_b$ across $m$ regions in a particular area given by set $\mc J$. For player $i\in\{a,b\}$, let $x^i_j\geq0$ denote the part of the fleet player $i$ assigned to region $j\in\mc J$ and $x^{-i}_j\geq0$ denote the part that its opponent assigned to the same region. Each player's decision vector can be defined by $x^i\defineas \text{col}(x^i_j)_{j\in\mc J}\in\Omega_i$, with the set of corresponding feasible allocations given by 
\begin{equation}
\label{eq:0}
   \Omega_i\defineas\{x^i\in\R^m_{\geq0} | \mathbf{1}_m^Tx^i=\mathbf{X}_i \land x^i_j\geq0, \forall j\in\mc J\}\,.
\end{equation}

Each company aims to split the ride-hailing fleet in an attempt to maximize the profit obtained by operating and recharging the vehicles in different regions. For a particular choice of strategies $x^a\in\Omega_a$ and $x^b\in\Omega_b$, we model the profit of each company as 
\begin{equation}
\label{eq:1}
    \mathbf{u}_i(x^i,x^{-i})\defineas\sum_{j\in \mc J}\mathbf{u}_{i,j}^{\text{market}}(x^i_j,x^{-i}_j)-\mathbf{u}_{i,j}^{\text{charge}}(x^i_j)\,,
\end{equation}
where $\mathbf{u}_{i,j}^{\text{market}}(x^i_j,x^{-i}_j)$ represents the expected ride-hailing market share the company can secure in region $j$ and $\mathbf{u}_{i,j}^{\text{charge}}(x^i_j)$ denotes the average cost of charging $x^i_j$ vehicles. 

\textbf{Market share model:} We assume the demand in the area is heterogeneously distributed between the regions, while the customers do not exhibit a preference for any particular company. In the demand-attractive regions, we anticipate a higher number of requests, necessitating a greater number of service vehicles to prevent passenger abandonment due to extended waiting times. In other words, we assume a portion of the revenue generated from ride-hailing requests will consistently be forfeited because passengers tend to cancel their requests if they wait too long for a vehicle to be assigned to them. Hence, we adopt the following model
\begin{equation}
\label{eq:2}
    \mathbf{u}_{i,j}^{\text{market}}(x^i_j,x^{-i}_j)\defineas N_jp_j\frac{x^i_j}{x^i_j+x^{-i}_j+\varepsilon_j}\,,
\end{equation}
where $N_j\in\R_{>0}$ denotes the average number of requests in the region $j$, $p_j \in \R_{>0}$ denotes the average profit per vehicle, and $\varepsilon_j>0$ models the profit loss due to abandonments. Therefore, the aggregated loss in area $j \in \mc J$ is given by
\begin{equation}
    \mathbf{p}^\text{loss}_j(x^a_j,x^{b}_j)\defineas N_jp_j\frac{\varepsilon_j}{x^a_j+x^{b}_j+\varepsilon_j} \,.
\end{equation}
By looking at $\mathbf{p}^\text{loss}_j$, it is clear that a higher number of service requests leads to a higher aggregate profit loss. Moreover, since $\mathbf{p}^\text{loss}_j$ is an increasing function of the abandonment parameter $\varepsilon_j$, a higher value of $\varepsilon_j$ will require more vehicles in the region to avoid large profit losses.

\textbf{Charging cost:} We assume that regions also offer different charging prices. Namely, in an attempt to discourage an excessive concentration of idle drivers in high-demand regions, we anticipate that less attractive regions would offer reduced charging prices. We adopt the following charging cost model
\begin{equation}
\label{eq:3}
    \mathbf{u}_{i,j}^{\text{charge}}(x^i_j)\defineas c_j\overline{d}_jx^i_j\,.
\end{equation}
Here, we define $c_j\in \R$ as the price per unit of energy and $\overline{d}_j\in \R$ as the average charging demand per vehicle.  In this context, we consider $\overline{d}_j$ to be influenced by the vehicle's operating time, the region's mean vehicle speed, and the battery discharge rate. Assuming that vehicles exhibit similar discharge rates and that idle vehicles continue to move in search of passengers, it is justified to regard this parameter as company-independent. 
\medskip

\noindent If $\beta^m_j\defineas N_jp_j$ and $\beta^c_j\defineas c_j\overline{d}_j$, then the interactions between the companies can be modeled by a set of two coupled optimization problems given by
\begin{equation}\label{eq:4}
    \mc G=\biggl\{\max_{x^i\in\Omega_i}\sum_{j\in \mc J}x^i_j\biggl(\frac{\beta^m_j}{x^i_j+x^{-i}_j+\varepsilon_j}-\beta^c_j\biggl),\forall i\in\{a,b\}\biggr\}.
\end{equation}
A viable solution to this game is the concept of a Nash equilibrium (NE). If we let $x\defineas\text{col}(x^i)_{i\in\{a,b\}}\in\Omega$, with $\Omega\defineas\Omega_a\times\Omega_b$, denote the joint strategy of the two companies, then the NE can be formally introduced as in Definition~\ref{def:1}.
\begin{definition}[Nash equilibrium]\label{def:1}
    A joint strategy $\overline{x}\in\Omega$ is a Nash equilibrium (NE) of the game $\mc G$ given by~\eqref{eq:4}, if for all $i\in\{a,b\}$ and all $x^i\in\Omega_i$ it holds that 
    \begin{equation}
        \label{eq:5}
        \mathbf{u}_i(\overline{x}^i,\overline{x}^{-i})\geq\mathbf{u}_i(x^i,\overline{x}^{-i})\,.
    \end{equation}
\end{definition}
\medskip

Although~\eqref{eq:4} resembles the standard lottery Blotto games with Tullock contest success functions~\cite{OSORIO2013164, KIM20181, LI2022102771}, there exist significant differences between the models. First, the proposed one takes into account that the total potential profit in each region will increase with the participation, while classical Tullock contests assume that the whole profit is split between the players, i.e., $\varepsilon_j = 0$. Second, the proposed model also includes charging costs in the form of linear terms, rendering the existing results on how to compute the NE in games with Tullock contests inapplicable.
Therefore, in the following section, we first prove the existence and uniqueness of the Nash equilibrium for the game $\mc G$ and then proceed to provide its general characterization. 

\section{Characterizing the Nash equilibria}\label{sec:NE}
Based on the characteristics of the cost functions and feasible sets outlined in Section~\ref{sec:model}, we first show that there exists a unique NE of the game $\mc G$ defined in~\eqref{eq:4}. 
\begin{theorem}
    Game $\mc G$ defined by~\eqref{eq:4}, with feasible sets $\Omega_i$ given by~\eqref{eq:0}, admits a unique Nash equilibrium.
\end{theorem}
\iftoggle{full_version}{\medskip\begin{proof}
    We start by observing that~\eqref{eq:1} is continuous in $x\in\Omega$ and concave in $x^i\in\Omega_i$. Indeed, for any given $x^{-i}$ we have
    \begin{equation}
        \frac{\partial^2\mathbf{u}_i}{\partial (x^i_j)^2}=\frac{-2\beta^m_j(x_j^{-i}+\varepsilon_j)}{(x^i_j+x^{-i}_j+\varepsilon_j)^3}<0\land\frac{\partial^2\mathbf{u}_i}{\partial x^i_j\partial x^i_k}=0\text{ if }j\neq k\,,
    \end{equation}
    which yields a negative definite diagonal Hessian $\nabla^2\mathbf{u}_i$. Next, by observing that the sets $\Omega_i$ are compact and convex, we can directly invoke~\cite[Th.1]{Rosen} to prove existence. Moreover, let us define
\begin{equation}
    \label{eq:38}
    g(x,r)\defineas\text{col}\left((r_{i}\nabla_{x^{i}}\mathbf{u}_i(x^i,x^{-i}))_{i\in \{a,b\}}\right)\,,
\end{equation}  
where $r=\text{col}((r_i)_{i\in\{a,b\}})\in\R^2_{>0}$. Let $G(x,r)$ denote the Jacobian of $g(x,r)$ with respect to $x$. A sufficient condition for the uniqueness of the Nash equilibrium is then that the matrix $\Gamma\defineas G(x,r)+G^{T}(x,r)$
be negative definite for all $x\in\Omega$ and some $r\in \R^2_{>0}$~\cite[Th.2]{Rosen}. For $r=\mathbf{1}_2$, we have
\begin{equation}
    \label{eq:39}
    \Gamma\defineas\left[\begin{array}{cc}
         2\mathbf{M}_{aa}& \mathbf{M}_s \\
         \mathbf{M}_s&2\mathbf{M}_{bb} 
    \end{array}\right]\,,
\end{equation}
where $\mathbf{M}_{aa}$, $\mathbf{M}_{bb}$ and $\mathbf{M}_s$ are given by
\begin{equation}
    \label{eq:40}   \mathbf{M}_{aa}=\text{diag}\left(\text{col}\left(\frac{-2\beta^m_j(x^b_j+\varepsilon_j)}{(x^a_j+x^b_j+\varepsilon_j)^3}\right)_{j\in\mc J}\right)\,,
\end{equation}
\begin{equation}
    \label{eq:42}   \mathbf{M}_{bb}=\text{diag}\left(\text{col}\left(\frac{-2\beta^m_j(x^a_j+\varepsilon_j)}{(x^a_j+x^b_j+\varepsilon_j)^3}\right)_{j\in\mc J}\right)\,,
\end{equation}
\begin{equation}
    \label{eq:43}   \mathbf{M}_{s}=\text{diag}\left(\text{col}\left(\frac{-2\beta^m_j\varepsilon_j}{(x^a_j+x^b_j+\varepsilon_j)^3}\right)_{j\in\mc J}\right)\,.
\end{equation}
By utilizing the properties of the Shur complement~\cite[p.34]{Schur}, $\Gamma\prec0$ is equivalent to $\mathbf{M}_{aa}\prec0$ and $\text{Sh}(2\mathbf{M}_{aa})\prec0$, where $\text{Sh}(2\mathbf{M}_{aa})\defineas 2\mathbf{M}_{bb}-\frac{1}{2}\mathbf{M}_{s}\mathbf{M}_{aa}^{-1}\mathbf{M}_{s}$ is the Schur complement of $2\mathbf{M}_{aa}$ in $\Gamma$. We calculate directly
\begin{equation}
    \label{eq:44}
    \text{Sh}(2\mathbf{M}_{aa})=\text{diag}\biggl(\text{col}\biggl(-\beta^m_j\frac{4x^a_j+\varepsilon_j(4-\frac{\varepsilon_j}{\varepsilon_j+x^b_j})}{(x^a_j+x^b_j+\varepsilon_j)^3}\biggr)_{j\in\mc J}\biggr)\,
\end{equation}
so since $\varepsilon_j>0$, we get$\frac{\varepsilon_j}{\varepsilon_j+x^b_j}\leq 1$, and hence $\text{Sh}(2\mathbf{M}_{aa})\prec0$. This implies $\Gamma \prec 0$ and hence the NE is unique.     
\end{proof}}{

The proof is given in the extended version of our paper.}

\subsection{General characterization}
Having established the uniqueness of the Nash equilibrium, to characterize it, we look at the best-response optimization problem of each player. Namely, for $\overline{x}\in\Omega$ to be a NE of~\eqref{eq:4}, for each $i\in\{a,b\}$, the strategy $\overline{x}^i\in\Omega_i$ has to solve the following best-response optimization problem 
\begin{maxi!}
    {x^i}{\sum_{j\in \mc J}x^i_j\biggl(\frac{\beta^m_j}{x^i_j+\overline{x}^{-i}_j+\varepsilon_j}-\beta^c_j\biggr)}
    {\label{maxi:op1}}{}
    \addConstraint{\mathbf{1}_m^Tx^i=\mathbf{X}_i}{}{\label{eq:cc1}}
    \addConstraint{x^i_j\geq 0 \,.}{}{\label{eq:cc2}}
    \end{maxi!}
Since the best-response optimization problem is concave, and $\Omega_i$ given by~\eqref{eq:cc1} and~\eqref{eq:cc2} is compact and convex, to characterize the Nash equilibrium it suffices to look at its set of KKT conditions. Let $\mc L$ be the Lagrangian of~\eqref{maxi:op1}
\begin{equation}
    \label{eq:6}
    \mc L_i(x^i)=\mathbf{u}_i(x^i,\overline{x}^{-i})+\lambda^i(\mathbf{1}^Tx^i-\mathbf{X}_i)+\sum_{j\in\mc J} \nu^i_jx^i_j\,,
\end{equation}
where $\lambda^i\in\R$ and $\nu^i_j\in\R_{\geq0}$ represent the dual variables associated with the one equality and $m$ inequality constraints of each player. For $\overline{x}\in\Omega$ to be a NE of~\eqref{eq:4}, it has to solve the set of nonlinear KKT equations of~\eqref{maxi:op1} for some feasible $\overline{\lambda}^i$ and $\overline{\nu}^i_j$. We start by analyzing the stationarity conditions
\begin{equation}
    \label{eq:7}
    \frac{\partial \mc L_a}{\partial x^a_j}=\overline{\lambda}^a+\overline{\nu}^a_j-\beta^c_j+\frac{\beta^m_j(\overline{x}^b_j+\varepsilon_j)}{(\overline{x}^a_j+\overline{x}^b_j+\varepsilon_j)^2}=0\,,
\end{equation}
\begin{equation}
    \label{eq:8}
    \frac{\partial \mc L_b}{\partial x^b_j}=\overline{\lambda}^b+\overline{\nu}^b_j-\beta^c_j+\frac{\beta^m_j(\overline{x}^a_j+\varepsilon_j)}{(\overline{x}^a_j+\overline{x}^b_j+\varepsilon_j)^2}=0\,,
\end{equation}
for all $j\in\mc J$. After rearranging~\eqref{eq:7} and~\eqref{eq:8}, we obtain
\begin{equation}
    \label{eq:9}
    \overline{x}^a_j=\frac{1}{\beta^m_j}(\beta^c_j-\overline{\lambda}^b-\overline{\nu}^b_j)(\overline{x}^a_j+\overline{x}^b_j+\varepsilon_j)^2-\varepsilon_j\,,
\end{equation}
\begin{equation}
    \label{eq:10}
    \overline{x}^b_j=\frac{1}{\beta^m_j}(\beta^c_j-\overline{\lambda}^a-\overline{\nu}^a_j)(\overline{x}^a_j+\overline{x}^b_j+\varepsilon_j)^2-\varepsilon_j\,.
\end{equation}
Let $\overline{t}_j\defineas \overline{x}^a_j+\overline{x}^b_j+\varepsilon_j$ and $\Delta_j\defineas 2\beta^c_j-\overline{\lambda}^a-\overline{\nu}^a_j-\overline{\lambda}^b-\overline{\nu}^b_j$. By combining~\eqref{eq:9} and~\eqref{eq:10}, we obtain a quadratic equation 
\begin{equation}
    \label{eq:11}
    \Delta_j\overline{t}_j^2-\beta_j^m\overline{t}_j-\beta_j^m\varepsilon_j=0\,.
\end{equation}
Because $\overline{x}^a_j,\overline{x}^b_j\geq0$ and $\varepsilon_j>0$,~\eqref{eq:11} yields the only viable option which requires that $\Delta_j>0$ and
\begin{equation}
    \label{eq:12}
\overline{x}^a_j+\overline{x}^b_j+\varepsilon_j=\frac{\beta_j^m+\sqrt{(\beta_j^m)^2+4\beta_j^m\varepsilon_j\Delta_j}}{2\Delta_j}\,.
\end{equation} 
Let $\mathbf{\Sigma}_{\varepsilon}\defineas\sum_{j\in\mc J}\varepsilon_j$, $\alpha_j\defineas2\beta^c_j-\overline{\nu}^a_j-\overline{\nu}^b_j$, and $\mathbf{t}_{\lambda}\defineas\overline{\lambda}^a+\overline{\lambda}^b$. By summing~\eqref{eq:12} over all regions $j\in\mc J$ we get
\begin{equation}
    \label{eq:13}
    \mathbf{X}_a+\mathbf{X}_b+\mathbf{\Sigma}_{\varepsilon}=\sum_{j\in\mc J}\frac{\beta_j^m+\sqrt{(\beta_j^m)^2+4\beta_j^m\varepsilon_j(\alpha_j-\mathbf{t}_{\lambda})}}{2(\alpha_j-\mathbf{t}_{\lambda})}\,.
\end{equation}
Solving the nonlinear equation~\eqref{eq:13} for $\mathbf{t}_\lambda$ represents the backbone for characterizing the Nash equilibrium. Therefore, we first prove the following lemma.
\iftoggle{full_version}{\begin{figure}[t]
    \centering
    \begin{tikzpicture}
        \begin{axis}[
            xlabel={$\mathbf{t}_{\lambda}$},
            ylabel={$f\left(\mathbf{t}_{\lambda}\right)$},
            domain=-5:15,
            samples=200,
            axis lines=middle,
            grid,
            ymin=-15, ymax=15, 
            xtick={\empty},
            ytick={\empty},
        ]
            \addplot[blue,thick] {(1+(1+0.1*(3-x))^0.5)/(3-x)+(1+(1+0.1*(5-x))^0.5)/(5-x)+(1+(1+0.1*(7-x))^0.5)/(7-x)-3};
            \addplot[black,dashed]{-3};
            \addplot[red,mark=*] coordinates {(1.9022,0)};
            \addplot[black,mark=*] coordinates {(3,0)};
            \addplot[black,mark=*] coordinates {(5,0)};
            \addplot[black,mark=*] coordinates {(7,0)};   
            
        \end{axis}
        \node(n1) at (2.92,2.6) {$\tilde{\alpha}_1$};
        \node(n2) at (3.62,2.6) {$\tilde{\alpha}_2$};
        \node(n3) at (4.32,2.6) {$\tilde{\alpha}_3$};

        \node[red](n4) at (2.25,3.2) {$\mathbf{t}_{\lambda}^*$};
        \node[black, ](n5) at (0.5,2) {\small$-\mathbf{X}_a-\mathbf{X}_b-\mathbf{\Sigma}_{\varepsilon}$};

        \fill[black, fill opacity=0.2] (6.15,0) rectangle (7,6);
        \draw[](6.15,0) -- (6.15,6);

    \end{tikzpicture}
    \caption{Illustrative example plot of the function~\eqref{eq:14} for $j\in\{1,2,3\}.$}
    \label{fig:1}
\end{figure}
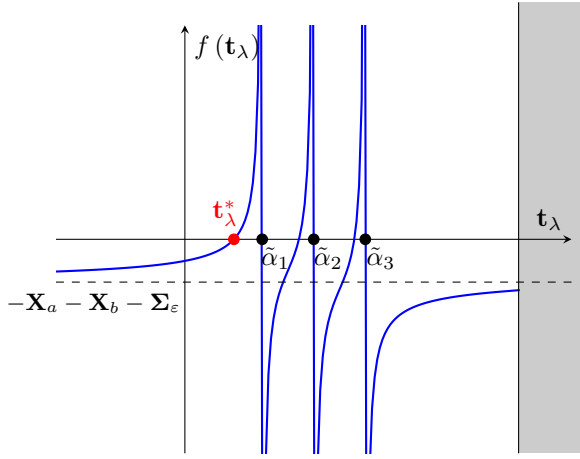}{}
\begin{lemma}\label{lm:3}
    Let the best-response optimization problem of each player $i\in\{a,b\}$ be defined by~\eqref{maxi:op1}. Then, for any two sets $\{\overline{\nu}^a_j\}_{j\in\mc J}$ and $\{\overline{\nu}^b_j\}_{j\in\mc J}$ such that
    $\overline{\nu}^a_j,\overline{\nu}^b_j\geq0$, there exist at most one $\overline{x}^a\in\Omega_a$ and $\overline{x}^b\in\Omega_b$ satisfying~\eqref{eq:7} and~\eqref{eq:8}.
\end{lemma}
\iftoggle{full_version}{\medskip
\begin{proof}
    First of all, let us define a function $f(\mathbf{t}_\lambda):\R\rightarrow\R$
    \begin{multline}
        \label{eq:14}
        f(\mathbf{t}_\lambda)=\sum_{j\in\mc J}\frac{\beta_j^m+\sqrt{(\beta_j^m)^2+4\beta_j^m\varepsilon_j(\alpha_j-\mathbf{t}_{\lambda})}}{2(\alpha_j-\mathbf{t}_{\lambda})} \\ -\mathbf{X}_a-\mathbf{X}_b-\mathbf{\Sigma}_{\varepsilon}\,.  
    \end{multline}
    It is clear that solutions of~\eqref{eq:13} correspond to zeros of $f(\mathbf{t}_\lambda)$ and that its domain is given by $(-\infty, \min_j\frac{\beta_j^m}{4\varepsilon_j}+\alpha_j]$. Observe that $f(\mathbf{t}_\lambda)$ has a vertical asymptote at $\alpha_j$ for all $j\in\mc J$ since $\lim_{\mathbf{t}_\lambda\rightarrow\alpha_j^+}f(\mathbf{t}_\lambda)=-\infty$ and $\lim_{\mathbf{t}_\lambda\rightarrow\alpha_j^-}f(\mathbf{t}_\lambda)=\infty$. Moreover, $\lim_{\mathbf{t}_\lambda\rightarrow-\infty}f(\mathbf{t}_\lambda)=-\mathbf{X}_a-\mathbf{X}_b-\mathbf{\Sigma}_{\varepsilon}$. If $\tilde{\alpha}_1$, $\tilde{\alpha}_2\hdots\tilde{\alpha}_m$ represent a non-decreasing permutation of $\alpha_j$ for all $j\in\mc J$, then Bolzano's theorem~\cite[Th.1]{Cauchy} ensures that there exists at least one zero of $f(\mathbf{t}_\lambda)$ in each of the intervals $(0,\tilde{\alpha}_1)$ and $(\tilde{\alpha}_k,\tilde{\alpha}_{k+1})$ for $1\leq k\leq m-1$. An illustration plot of $f(\mathbf{t}_\lambda)$ for $j\in\{1,2,3\}$ is shown in Figure~\ref{fig:1}. However, for any $\overline{\nu}^a_j,\overline{\nu}^b_j\geq0$,~\eqref{eq:11} implies $\Delta_j>0$ for all $j\in\mc J$. This means that all feasible solutions to~\eqref{eq:13} satisfy $\mathbf{t}_\lambda^*<\min_{j\in\mc J}\alpha_j=\tilde{\alpha}_1$. Hence, all the feasible solutions must be in $(0,\tilde{\alpha}_1)$. Looking at the first derivative of $f(\mathbf{t}_\lambda)$
    \begin{equation}
        \label{eq:15}
        f'(\mathbf{t}_\lambda)=\sum_{j\in\mc J}\frac{\beta_j^m}{2(\alpha_j-\mathbf{t}_\lambda)^2}\left(1+\frac{1+2\frac{\varepsilon_j}{\beta_j^m}(\alpha_j-\mathbf{t}_\lambda)}{\sqrt{1+4\frac{\varepsilon_j}{\beta_j^m}(\alpha_j-\mathbf{t}_{\lambda})}}\right)\,,
    \end{equation}
    we see that $f'(\mathbf{t}_\lambda)>0$ on $(0,\tilde{\alpha}_1)$, meaning that $f(\mathbf{t}_\lambda)$ is strictly increasing on this interval. In return, $f(\mathbf{t}_\lambda)$ can have at most one zero on $(0,\tilde{\alpha}_1)$, which combined with Bolzano's theorem guarantees a unique $\mathbf{t}_\lambda^*$. By substituting $\mathbf{t}_\lambda^*$ into~\eqref{eq:12}, we can define $\overline{x}^a_j+\overline{x}^b_j+\varepsilon_j=\kappa_j^*$ and hence 
    \begin{equation}
        \label{eq:16}
        \overline{x}^i_j=\frac{(\kappa_j^*)^2}{\beta_j^m}(\beta_j^c-\overline{\lambda}^{-i}-\overline{\nu}^{-i}_j)-\varepsilon_j\,,
    \end{equation}
    holds for $i\in\{a,b\}$. By summing~\eqref{eq:16} for all $j\in\mc J$ we get
    \begin{multline}\label{eq:17}
        \overline{\lambda}^{i}= \\ \frac{1}{\sum_j\frac{(\kappa_j^*)^2}{\beta^m_j}}\biggl(-\mathbf{\Sigma}_\varepsilon-\mathbf{X}_{-i}+\sum_j\frac{\beta_j^c(\kappa^*_j)^2}{\beta^m_j}-\sum_j\frac{\overline{\nu}^i_j(\kappa^*_j)^2}{\beta^m_j}\biggr)
        \end{multline}
    which uniquely determines the multipliers $\overline{\lambda}^a$ and $\overline{\lambda}^b$. Plugging $\mathbf{t}_\lambda^*$ and~\eqref{eq:17} into~\eqref{eq:9} and~\eqref{eq:10} directly yields unique values of $\overline{x}^a$ and $\overline{x}^b$. If now $\overline{x}^a_j,\overline{x}^b_j\geq0$ then there exists a unique solution, otherwise, for this choice of $\overline{\nu}_j^a$ and $\overline{\nu}_j^b$ there exists no feasible solution.  
\end{proof}
\medskip}{

The proof is given in the extended version of our paper.\medskip}

Observe that $(\overline{x}^a,\overline{x}^b, \{\overline{\nu}^a_j\}_{j\in\mc J},\{\overline{\nu}^b_j\}_{j\in\mc J}, \overline{\lambda}^a,\overline{\lambda}^b)$ determined by~\eqref{eq:9} and~\eqref{eq:10} will be a solution of the best-response optimization problem~\eqref{maxi:op1}, if the complementary slackness condition also holds for all $j\in\mc J$. Namely, if $\overline{\nu}^i_j>0$ for some $i,j$, then it has to hold that $\overline{x}_j^i=0$. Since $\mathbf{t}_\lambda^*$ has to be computed numerically and the choice of $\overline{\nu}_j^i$ directly influences the interval in which $\mathbf{t}_\lambda^*$ is located, finding the NE of~\eqref{eq:4} for $m>2$ is in general hard as it 
requires extensive exploration of the unbounded space of dual variables $\overline{\nu}^i_j$. However, if the unique NE of~\eqref{eq:4} is in the interior of the feasible set, i.e., it is a strategy $\overline{x}\in\Omega$ with $\overline{x}_j^a,\overline{x}^b_j>0$ for all $j\in\mc J$, then Theorem~\ref{th:2} directly allows us to compute it.
\begin{theorem}\label{th:2}
    Let the best-response optimization problem of each player $i\in\{a,b\}$ in game~\eqref{eq:4} be defined by~\eqref{maxi:op1}. Then, if the interior NE strategy $\overline{x}\in\Omega$ exists, it is determined by
    \begin{equation}
        \overline{x}^i_j=\frac{(\kappa_j^*)^2}{\beta_j^m}\left(\beta_j^c-\overline{\lambda}^{-i}\right)-\varepsilon_j\,,
    \end{equation}
    \begin{equation}
        \overline{\lambda}^{i}=\frac{1}{\sum_j\frac{(\kappa_j^*)^2}{\beta^m_j}}\biggl(-\mathbf{\Sigma}_\varepsilon-\mathbf{X}_{-i}+\sum_j\frac{\beta_j^c(\kappa^*_j)^2}{\beta^m_j}\biggr)\,,
    \end{equation}
    where for all $j\in\mc J$, $\kappa_j^*$ is given by
    \begin{equation}
        \kappa_j^*=\frac{\beta_j^m+\sqrt{(\beta_j^m)^2+4\beta_j^m\varepsilon_j(2\beta^c_j-\mathbf{t}_\lambda^*)}}{2(2\beta^c_j-\mathbf{t}_\lambda^*)}\,,
    \end{equation}
    where $\mathbf{t}_\lambda^*$ is the unique solution of~\eqref{eq:13} with $\alpha_j=2\beta^c_j$ and satisfying $\mathbf{t}_\lambda^*\in(-\infty,\min_j \frac{\beta^m_j}{4\varepsilon_j}+2\beta^c_j]$.

\end{theorem}
\medskip
On the other hand, when there are only two regions, i.e., $\mc J=\{1,2\}$, apart from checking for the interior NE that can be computed using Theorem~\ref{th:2}, we can also analytically examine the existence of a Nash equilibrium on the boundary of the feasible set. Therefore, for a two-region setup, we can always compute the unique NE. 
\subsection{Two-region case}
The number of local regions in a given area is essentially determined by the range of charging prices accessible for vehicles to choose from and the desire to have homogeneously distributed demand within the region. Therefore, it is not unlikely that a city area could be divided into a downtown and a suburban region such that the vehicles incur lower expenses should they choose to charge in the suburban areas. Interestingly, this setup allows us to establish additional criteria based on $\beta^m_j$, $\beta^c_j$, $\varepsilon_j$, $\mathbf{X}_a$ and $\mathbf{X}_b$ that precisely locate the NE. 

Due to complementary slackness, characterizing the Nash equilibria on the boundary in this case boils down to examining if some of the following structures
\begin{itemize}
    \item $\overline{x}^T=[0, \mathbf{X}_a, z, \mathbf{X}_b-z]$ for $z\in[0,\mathbf{X}_b]$,
    \item $\overline{x}^T=[\mathbf{X}_a, 0, z, \mathbf{X}_b-z]$ for $z\in[0,\mathbf{X}_b]$,
    \item $\overline{x}^T=[z, \mathbf{X}_a-z, 0, \mathbf{X}_b]$ for $z\in[0,\mathbf{X}_a]$,
    \item $\overline{x}^T=[z, \mathbf{X}_a-z, \mathbf{X}_b, 0]$ for $z\in[0,\mathbf{X}_a]$,
\end{itemize}
can represent a NE of the game~\eqref{eq:4}, i.e., if it can be a solution to a pair of best-response optimization problems~\eqref{maxi:op1}. Given the symmetry of the problems concerning the four cases mentioned above, in the two subsequent lemmas, we will present our findings for scenarios in which player $a$ chooses a boundary action, i.e., the first two cases. The analysis for the remaining two cases is completely analogous.
\begin{lemma}\label{lm:1}
     Let the best-response optimization problem of each player $i\in\{a,b\}$ in game~\eqref{eq:4} be defined by~\eqref{maxi:op1} and $\mc J=\{1,2\}$. Furthermore, let us define $\mathbf{\overline{V}}>\mathbf{\underline{V}}$ as
     \begin{equation}
         \label{eq:18}
        \mathbf{\overline{V}}\defineas\beta^c_2-\beta^c_1+\frac{\beta^m_1}{\varepsilon_1}-\frac{\beta^m_2(\mathbf{X}_a+\varepsilon_2)}{(\mathbf{X}_a+\mathbf{X}_b+\varepsilon_2)^2}\,,
     \end{equation}
    \begin{equation}
         \label{eq:19}
        \mathbf{\underline{V}}\defineas\beta^c_2-\beta^c_1+\frac{\beta^m_1\varepsilon_1}{(\mathbf{X}_b+\varepsilon_1)^2}-\frac{\beta^m_2}{\mathbf{X}_a+\varepsilon_2}\,.
     \end{equation}
     If $\overline{x}^T=[0, \mathbf{X}_a, z^*, \mathbf{X}_b-z^*]$ solves~\eqref{maxi:op1} for player $b$, then $z^*=\mathbf{X}_b$ if $\mathbf{\underline{V}}\geq0$, $z^*=0$ if $\mathbf{\overline{V}}\leq0$ and in all other cases, $z^*$ represents the unique solution of the equation 
    \begin{equation}
        \label{eq:20}
        \frac{\beta^m_1\varepsilon_1}{(z+\varepsilon_1)^2}-\frac{\beta^m_2(\mathbf{X}_a+\varepsilon_2)}{(\mathbf{X}_a+\mathbf{X}_b-z+\varepsilon_2)^2}+\beta^c_2-\beta^c_1=0
    \end{equation}
    on interval $[0,\mathbf{X}_b]$.
\end{lemma}
\iftoggle{full_version}{\medskip\begin{proof} 
We start by recalling that the best-response optimization of player $b$ 
    is concave for $\overline{x}^a=[0,\mathbf{X}_a]^T$, hence $\frac{\partial^2\mathbf{u}_b}{\partial z^2}<0$. This means that the first derivative $\frac{\partial\mathbf{u}_b}{\partial z}$ given by
    \begin{equation}
        \label{eq:21}
        \frac{\partial\mathbf{u}_b}{\partial z}=\frac{\beta^m_1\varepsilon_1}{(z+\varepsilon_1)^2}-\frac{\beta^m_2(\mathbf{X}_a+\varepsilon_2)}{(\mathbf{X}_a+\mathbf{X}_b-z+\varepsilon_2)^2}+\beta^c_2-\beta^c_1\,,
    \end{equation}
    is decreasing in $z$ on the interval $[0,\mathbf{X}_b]$. Therefore, the maximum and minimum values of $\frac{\partial\mathbf{u}_b}{\partial z}$ are attained for $z=0$ and $z=\mathbf{X}_b$ respectively, i.e., 
    \begin{equation}
        \label{eq:22}
            \left.\frac{\partial\mathbf{u}_b}{\partial z}\right|_{z=0}=\mathbf{\overline{V}}\land\left.\frac{\partial\mathbf{u}_b}{\partial z}\right|_{z=\mathbf{X}_b}=\mathbf{\underline{V}}\,.
    \end{equation}
    If $\mathbf{\underline{V}}\geq0$, then $\frac{\partial\mathbf{u}_b}{\partial z}\geq0$. Consequently, $\mathbf{u}_b$ is increasing for $z\in[0,\mathbf{X}_b]$. Since the best-response optimization problem concerns maximization of $\mathbf{u}_b$, the optimal response is $z^*=\mathbf{X}_b$. On the other hand, if $\mathbf{\overline{V}}\leq0$, then $\frac{\partial\mathbf{u}_b}{\partial z}\leq0$ and $\mathbf{u}_b$ is decreasing for $z\in[0,\mathbf{X}_b]$. Therefore, the optimal response is $z^*=0$. Finally, if $\mathbf{\overline{V}}\mathbf{\underline{V}}<0$, then based on Bolzano's theorem~\cite[Th.1]{Cauchy} and the fact that $\frac{\partial\mathbf{u}_b}{\partial z}$ is monotone, there exists a unique $z^*\in[0,\mathbf{X}_b]$ such that $\left.\frac{\partial\mathbf{u}_b}{\partial z}\right|_{z=z^*}=0$. Because $\mathbf{u}_b$ is concave, the stationary point that solves~\eqref{eq:20} corresponds to the maximum of $\mathbf{u}_b$.
\end{proof}}{

The proof is given in the extended version of our paper.}
\medskip
\begin{lemma}\label{lm:2}
     Let the best-response optimization problem of each player $i\in\{a,b\}$ in game~\eqref{eq:4} be defined by~\eqref{maxi:op1} and $\mc J=\{1,2\}$. Furthermore, let us define $\mathbf{\overline{W}}>\mathbf{\underline{W}}$ as
     \begin{equation}
         \label{eq:28}
        \mathbf{\overline{W}}\defineas\beta^c_2-\beta^c_1+\frac{\beta^m_1}{\mathbf{X}_a+\varepsilon_1}-\frac{\beta^m_2\varepsilon_2}{(\mathbf{X}_b+\varepsilon_2)^2}\,,
     \end{equation}
    \begin{equation}
         \label{eq:29}
        \mathbf{\underline{W}}\defineas\beta^c_2-\beta^c_1+\frac{\beta^m_1(\mathbf{X}_a+\varepsilon_1)}{(\mathbf{X}_a+\mathbf{X}_b+\varepsilon_1)^2}-\frac{\beta^m_2}{\varepsilon_2}\,.
     \end{equation}
     If $\overline{x}^T=[\mathbf{X}_a, 0, z^*, \mathbf{X}_b-z^*]$ solves~\eqref{maxi:op1} for player $b$, then $z^*=\mathbf{X}_b$ if $\mathbf{\underline{W}}\geq0$, $z^*=0$ if $\mathbf{\overline{W}}\leq0$ and in all other cases, $z^*$ represents the unique solution of the equation 
    \begin{equation}
        \label{eq:30}
        \frac{\beta^m_1(\mathbf{X}_a+\varepsilon_1)}{(\mathbf{X}_a+z+\varepsilon_1)^2}-\frac{\beta^m_2\varepsilon_2}{(\mathbf{X}_b-z+\varepsilon_2)^2}+\beta^c_2-\beta^c_1=0
    \end{equation}
    on interval $[0,\mathbf{X}_b]$.
\end{lemma}
\iftoggle{full_version}{
\medskip
\begin{proof}
    Similar to the proof of Lemma~\ref{lm:1}, the first derivative $\frac{\partial\mathbf{u}_b}{\partial z}$, that is decreasing in $z$ for $z\in[0,\mathbf{X}_b]$, is given by
    \begin{equation}
        \label{eq:31}
        \frac{\partial\mathbf{u}_b}{\partial z}=\frac{\beta^m_1(\mathbf{X}_a+\varepsilon_1)}{(\mathbf{X}_a+z+\varepsilon_1)^2}-\frac{\beta^m_2\varepsilon_2}{(\mathbf{X}_b-z+\varepsilon_2)^2}+\beta^c_2-\beta^c_1\,,      
    \end{equation}
    and consequently satisfies
    \begin{equation}
        \label{eq:32}
            \left.\frac{\partial\mathbf{u}_b}{\partial z}\right|_{z=0}=\mathbf{\overline{W}}\land\left.\frac{\partial\mathbf{u}_b}{\partial z}\right|_{z=\mathbf{X}_b}=\mathbf{\underline{W}}\,.
    \end{equation}
    With the same reasoning as in the proof of Lemma~\ref{lm:1} we can now conclude the statement of Lemma~\ref{lm:2}.
\end{proof}}{

The proof is given in the extended version of our paper.}

\medskip
Using Lemmas~\ref{lm:1} and~\ref{lm:2}, we can now outline the conditions under which $\overline{x}^T=[0, \mathbf{X}_a, z^*, \mathbf{X}_b-z^*]$ and $\overline{x}^T=[\mathbf{X}_a, 0, z^*, \mathbf{X}_b-z^*]$, with $z^*\in[0,\mathbf{X}_b]$ chosen so as to solve the best-response optimization problem~\eqref{maxi:op1} of player $b$, represent the Nash equilibrium of game~\eqref{eq:4}.
\begin{theorem}\label{th:3}
    Let the best-response optimization problem of each player $i\in\{a,b\}$ in~\eqref{eq:4} be defined by~\eqref{maxi:op1} and $\mc J=\{1,2\}$. Let $z_1^*,z^*_2\in[0,\mathbf{X}_b]$ be such that $\overline{x}_1^T=[0, \mathbf{X}_a, z_1^*, \mathbf{X}_b-z_1^*]$ and $\overline{x}_2^T=[\mathbf{X}_a, 0, z_2^*, \mathbf{X}_b-z_2^*]$ solve the best-response optimization problem of player $b$. Then, $\overline{x}_1$ is the NE of game~\eqref{eq:4} if and only if 
    \begin{equation}
        \label{eq:23}
        z_1^*\in(0,\mathbf{X}_b)\:\land\: \frac{(\mathbf{X}_b-z_1^*-\mathbf{X}_a)\beta^m_2}{(\mathbf{X}_a+\mathbf{X}_b-z_1^*+\varepsilon_2)^2}-\frac{\beta^m_1z_1^*}{(z_1^*+\varepsilon_1)^2}\geq0\,,
    \end{equation}
    or
    \begin{equation}
        \label{eq:24}
        z_1^*=0\:\land\: -\mathbf{\overline{V}}+\frac{(\mathbf{X}_b-\mathbf{X}_a)\beta^m_2}{(\mathbf{X}_a+\mathbf{X}_b+\varepsilon_2)^2}\geq0\,.
    \end{equation}
    In addition, $\overline{x}_2$ is the NE of game~\eqref{eq:4} if and only if
    \begin{equation}
        \label{eq:33}
        z_2^*\in(0,\mathbf{X}_b)\:\land\: 
        \frac{(z_2^*-\mathbf{X}_a)\beta^m_1}{(\mathbf{X}_a+z_2^*+\varepsilon_1)^2}-\frac{(\mathbf{X}_b-z_2^*)\beta^m_2}{(\mathbf{X}_b-z_2^*+\varepsilon_2)^2}\geq0\,,
    \end{equation}
    or
    \begin{equation}
        \label{eq:34}
        z_2^*=\mathbf{X}_b\:\land\: \mathbf{\underline{W}}+\frac{(\mathbf{X}_b-\mathbf{X}_a)\beta^m_1}{(\mathbf{X}_a+\mathbf{X}_b+\varepsilon_1)^2}\geq0\,.
    \end{equation}
\end{theorem}
\iftoggle{full_version}{
\medskip
\begin{proof}
    For $\overline{x}_1$ to be a NE, we have to verify that $\overline{x}^a_1=[0,\mathbf{X}_a]$ is a solution to the best-response optimization problem of player $a$ for strategy $\overline{x}^b_1=[z_1^*,\mathbf{X}_b-z_1^*]$ of player $b$. Due to complementary slackness, we have that $\overline{\nu}^a_2=0$, so what is left is to verify if there exists a feasible $\overline{\nu}^a_1\geq0$. For $z_1^*\in(0,\mathbf{X}_b)$, plugging $\overline{x}^a_1$ and $\overline{x}_1^b$ into~\eqref{eq:7} and~\eqref{eq:8}, and then subtracting~\eqref{eq:8} from~\eqref{eq:7} with~\eqref{eq:20} in mind, yields
    \begin{equation}
        \label{eq:25}
        \overline{\nu}^a_1 = \frac{(\mathbf{X}_b-z_1^*-\mathbf{X}_a)\beta^m_2}{(\mathbf{X}_a+\mathbf{X}_b-z_1^*+\varepsilon_2)^2}-\frac{\beta^m_1z_1^*}{(z_1^*+\varepsilon_1)^2}\,,
    \end{equation}
    hence the first condition. Similarly, if $z_1^*=0$ then $\mathbf{\overline{V}}\leq0$ by Lemma~\ref{lm:1}, so the same steps as before result in
    \begin{equation}
        \label{eq:26}
        \overline{\nu}^a_1 =  -\mathbf{\overline{V}}+\frac{(\mathbf{X}_b-\mathbf{X}_a)\beta^m_2}{(\mathbf{X}_a+\mathbf{X}_b+\varepsilon_2)^2}\geq0\,,
    \end{equation}
    hence the second condition. Finally, if $z_1^*=\mathbf{X}_b$, then $\mathbf{\underline{V}}\geq0$ by Lemma~\ref{lm:1} and we get
    \begin{equation}
        \label{eq:27}
        \overline{\nu}^a_1 =  -\mathbf{\underline{V}} -\frac{\beta_1^m\mathbf{X}_b}{(\mathbf{X}_b+\varepsilon_1)^2}-\frac{\beta_2^m\mathbf{X}_a}{(\mathbf{X}_a+\varepsilon_2)^2}<0\,,
    \end{equation}
    which is not a feasible solution.
    Similarly, for $\overline{x}_2$, due to complementary slackness, we have $\overline{\nu}_1^a=0$. By plugging $\overline{x}^a_2=[\mathbf{X}_a,0]$ and $\overline{x}^b_2=[z_2^*,\mathbf{X}_b-z_2^*]$ into~\eqref{eq:7} and~\eqref{eq:8}, and then subtracting~\eqref{eq:7} from~\eqref{eq:8}, we obtain
    respectively
    \begin{equation}
        \label{eq:35}
        \overline{\nu}^a_2=\frac{(z_2^*-\mathbf{X}_a)\beta^m_1}{(\mathbf{X}_a+z_2^*+\varepsilon_1)^2}-\frac{(\mathbf{X}_b-z_2^*)\beta^m_2}{(\mathbf{X}_b-z_2^*+\varepsilon_2)^2}\quad\text{and}\,,
    \end{equation}
    \begin{equation}
        \label{eq:36}
        \overline{\nu}^a_2=\mathbf{\underline{W}}+\frac{(\mathbf{X}_b-\mathbf{X}_a)\beta^m_1}{(\mathbf{X}_a+\mathbf{X}_b+\varepsilon_1)^2}\,,
    \end{equation}
    for $z_2^*\in(0,\mathbf{X}_b)$ and $z_2^*=\mathbf{X}_b$. Finally, for $z_2^*=0$, based on Lemma~\ref{lm:2}, we have $\mathbf{\overline{W}}\leq0$ and hence
    \begin{equation}
        \label{eq:37}
        \overline{\nu}^a_2=\mathbf{\overline{W}}-\frac{\beta_1^m\mathbf{X}_a}{(\mathbf{X}_a+\varepsilon_1)^2}-\frac{\beta_2^m\mathbf{X}_b}{(\mathbf{X}_b+\varepsilon_2)^2}<0\,,
    \end{equation}
    cannot be a feasible solution to the best-response optimization problem of player $a$.
\end{proof}}{

The proof is given in the extended version of our paper.}
\medskip
\iftoggle{full_version}{\begin{remark}
    Interestingly, Theorem~\ref{th:3} reveals that in the Nash equilibrium, it is impossible for one company to exclusively serve one region while the other company exclusively serves the other region. However, there is a possibility that a particular area may remain completely unserved. 
\end{remark}
\medskip}{}

\noindent It is now clear that combining Theorem~\ref{th:2},~\ref{th:3}
and their counterparts corresponding to structures $\overline{x}^T=[z,\mathbf{X}_a-z,0,\mathbf{X}_b]$ and $\overline{x}^T=[z,\mathbf{X}_a-z, \mathbf{X}_b, 0]$, allows us to compute the Nash equilibrium of any two-region game~\eqref{eq:4}. In the following section, we will demonstrate the outlined procedures and explore the influence of varying the charging prices.   

\section{Numerical Example}\label{sec:example}
As outlined in Section~\ref{sec:model}, in this paper we are interested in analyzing the static scenarios assuming that vehicles assigned to a particular region stay there to charge. To demonstrate the performance of the algorithms, we construct a four-region and a two-region setup and analyze the attained NE.
\iftoggle{full_version}{\begin{figure}[!t]
    \begin{adjustbox}{ max width=0.48\textwidth}
    \centering
    \input{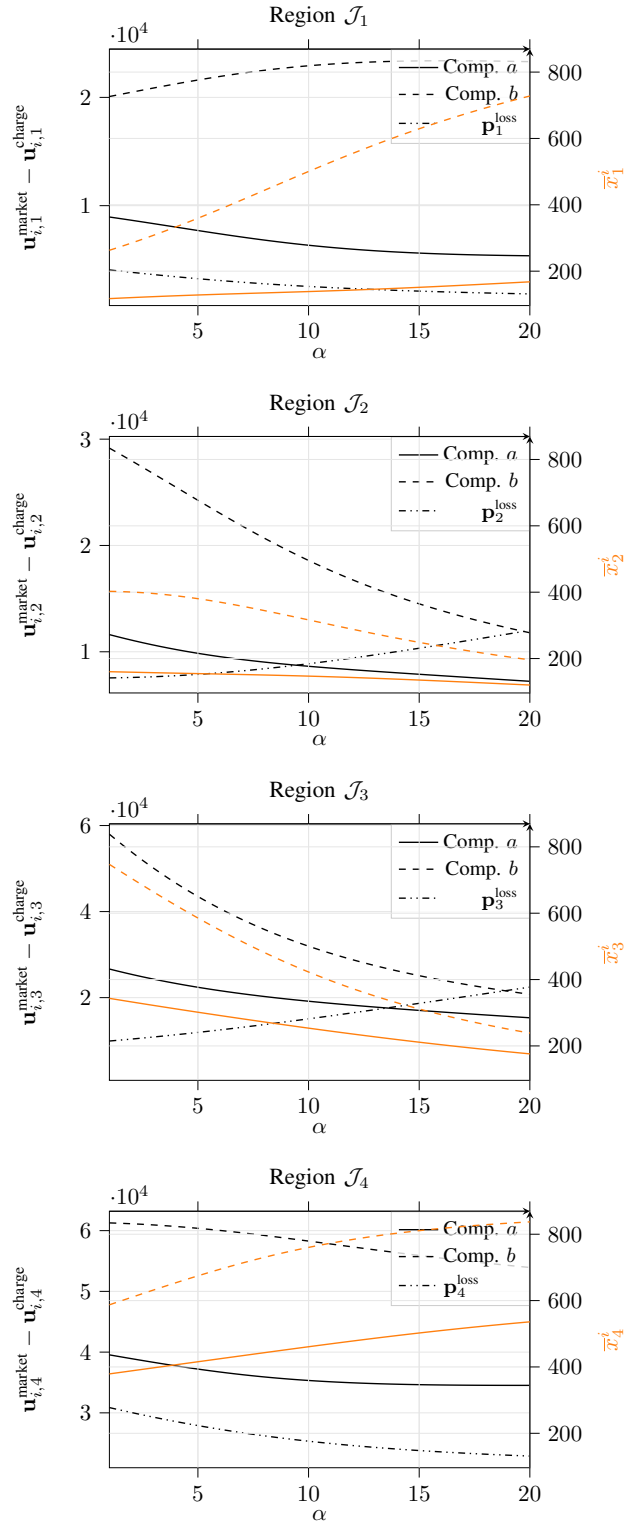}
    \end{adjustbox}
    \caption{\textbf{Four-region setup} - Fleet distribution and earned profit of the companies in the attained NE for different values of parameter $\alpha$.}
    \label{fig:alg1}
\end{figure}}{\begin{figure}[!t]
    \begin{adjustbox}{ max width=0.48\textwidth}
    \centering
    \input{figures/BigFigure.tikz}
    \end{adjustbox}
    \caption{\textbf{Four-region setup} - Fleet distribution and earned profit of the companies in the attained NE for different values of parameter $\alpha$.}\vspace{-1.5em}
    \label{fig:alg1}
\end{figure}}

\subsection{Four-region case}
We start by examining a scenario wherein the city is segmented into four regions, denoted as $\mc J=\{ J_1,  J_2,  J_3,  J_4\}$, each characterized by varying levels of attractiveness in terms of ride-hailing demand. In this scenario, we assume the companies operate with fleet sizes represented by $\mathbf{X}_a=1000$ and $\mathbf{X}_b=2000$ vehicles, respectively. The profit potential across these regions is depicted by the vector $\beta^m=[35\cdot 10^3, 50\cdot 10^3, 100\cdot 10^3, 180\cdot 10^3]^T$, while the levels of abandonments are detailed through the vector $\varepsilon=[50, 100, 120, 200]^T$. The setup indicates that $J_4$ exhibits the highest demand for ride-hailing services, necessitating the largest number of vehicles to serve it effectively. Conversely, $J_1$ reflects the lowest demand, resulting in a comparatively smaller number of abandoned service requests. Regarding the charging costs, we assume they are determined by a parametrized vector $\beta^c=[5, 3\alpha, 5\alpha, 50]^T$, with $\alpha\in[1,20]$. 

We simulate the model for $N_{\text{total}}=100$ values of parameter $\alpha$ in an attempt to illustrate the impact of the charging prices on the attained Nash equilibrium. For the chosen configuration of parameters for the four-region setup, we were able to find the unique Nash equilibrium by direct application of Theorem~\ref{th:2} for every value of $\alpha$. In Figure~\ref{fig:alg1}, we show the secured profits of each company in the NE and the corresponding fleet split for each $\alpha$. It is evident that the company operating a larger fleet is notably more affected by the increased charging prices. Additionally, we observe that in regions $J_2$ and $J_3$, as the parameter $\alpha$ increases, both companies simultaneously decrease the portion of their fleet. This reduction results in fewer vehicles available for service, leading to an increase in the incurred profit loss $\mathbf{p}^{\text{loss}}_j$. This is further supported by observing that regions $J_1$ and $J_4$ with fixed charging costs become increasingly popular as $\alpha$ increases. Furthermore, the plot also implies that the reduction in vehicle engagement is more prominent in $J_3$ compared to $J_2$ as $\alpha$ increases, which aligns with the fact that the charging prices rise more rapidly in $J_3$.   

\subsection{Two-region case}
To test the procedure for finding the NE irrespective of its position in the feasible set, we construct a two-region setup, i.e., $J=\{J_1,J_2\}$. In this case, the potential profit and abandonments are described by $\beta^m=[35\cdot 10^3,120\cdot 10^3]^T$ and $\varepsilon=[100,300]^T$ and we keep the same fleet sizes as for the four-region case. To model the charging costs we now adopt a parametrized vector $\beta^c=[10, 10\alpha]^T$ with $\alpha\in[1,50]$. Similar to the previous setup, we show the achieved fleet distribution and attained profits per region for different values of $\alpha$ in Figure~\ref{fig:alg2}. Apart from observing the same patterns as in the four-region case, here it is also worth noting that there exists a critical value of $\alpha$ beyond which the Nash equilibrium will be located on the boundary of the feasible set. 
\begin{table}
\begin{center}
 \renewcommand{\arraystretch}{1.3}
 \caption{Company decisions and attained profit}\vspace{1ex}
 \label{tab:res}
 \begin{tabular}{c|cc|c|cc|c}
 \multirow{2}{*}{$\alpha$} & \multicolumn{3}{c|}{Company $a$} & \multicolumn{3}{c}{Company $b$} \\ \cline{2-7}
 & $\overline{x}_{1}^{a}$ & $\overline{x}_{2}^{a}$ & $\mathbf{u}_a$ & $\overline{x}_{1}^{b}$ & $\overline{x}_{2}^{b}$ & $\mathbf{u}_b$ \\
 \hline
 $1.0 \rule{0pt}{2.6ex}$ &  222.6 & 777.4  & 35591 & 453.0  & 1547.0 & 71178.4   \\
 5.0 &  484.1 & 515.9  & 20448 & 1336.6  & 663.4 & 32165.6   \\
 25.0 &  943.6 & 56.4  & 3707.7 & 1937.9  & 62.1 & 5646.1  \\
 41.0 &  1000.0 & 0.0  & 1290.3 & 2000.0  & 0.0 & 2580.7  \\
\hline 
\end{tabular}
\end{center}
\end{table}
\iftoggle{full_version}{\begin{figure}[!t]
    \centering
    \begin{adjustbox}{ max width=0.48\textwidth}
    \input{figures/BigFigure1Ex.tikz}
    \end{adjustbox}
    \caption{\textbf{Two-region setup} - Attained NE of the companies and secured profits in different regions for different values of parameter $\alpha$. The vertical black line represents the experimentally obtained critical value $\alpha_{\text{crit}}$.}
    \label{fig:alg2}
\end{figure}}{\begin{figure}[!t]
    \centering
    \begin{adjustbox}{ max width=0.48\textwidth}
    \input{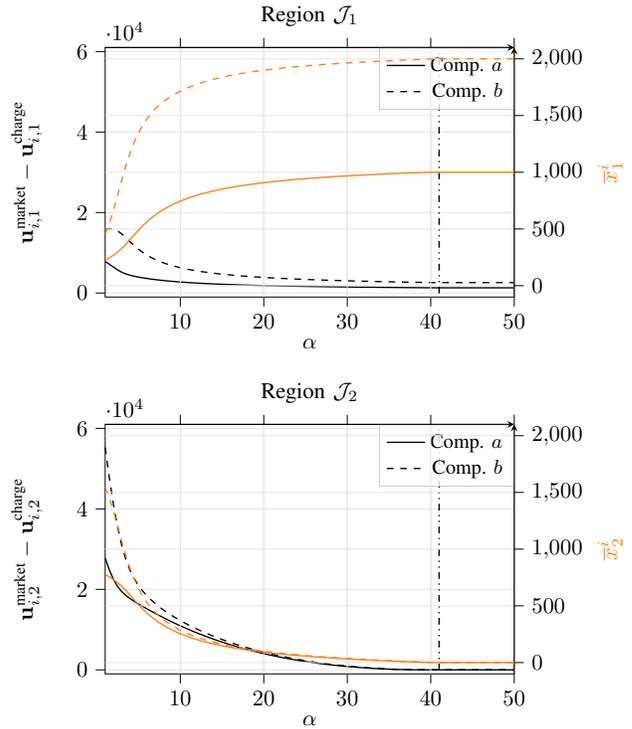}
    \end{adjustbox}
    \caption{\textbf{Two-region setup} - Attained NE of the companies and secured profits in different regions for different values of parameter $\alpha$. The vertical black line represents the experimentally obtained critical value $\alpha_{\text{crit}}$.}
    \label{fig:alg2}\vspace{-1.5em}
\end{figure}}
For the chosen set of parameters, by numerical inspection, we get the approximate value of $\alpha_{\text{crit}}\approx40.7$. For reference, in Table~\ref{tab:res}, we show the numerical values of the attained Nash equilibrium and the total achieved profit of the companies for some specific values of $\alpha$. As expected, with the increase in $\alpha$, the second region $J_2$ becomes less attractive even though it enjoys higher ride-hailing service demand. Beyond the critical value $\alpha_{\text{crit}}$, the Nash equilibrium of the system imposes that both companies send their whole fleets to serve region $J_1$ as the charging costs in region $J_2$ would become predominant. However, although theoretically possible, the high value of $\alpha_{\text{crit}}$ suggests that this setup would correspond to an unrealistic real-world scenario as it is highly unlikely that the charging prices in one region would be almost 41 times greater than in the other region. 

Finally, we explore the impacts of expanding the fleet size of one of the companies. We assume company $a$ remains with the same fleet of $\mathbf{X}_a=1000$ vehicles but we let $\mathbf{X}_b\in[200,4000]$. For the same $\beta^m$ and $\beta^c=[10.0,30.0]$, Figure~\ref{fig:alg3} shows the splits and the total profits of the companies. As expected, with the increase of $\mathbf{X}_b$ the total profit of company $a$ decreases due to its smaller market share. However, the plot also suggests that for a fixed fleet size of company $a$, there exists an optimal value of $\mathbf{X}_b$ resulting in the highest total profit. This is not unexpected since a bigger fleet also incurs higher charging costs thereby diminishing the advantages achieved when securing the market share. For the chosen parameters, by numerical inspection, we get $\mathbf{X}_b^{\text{optimal}}\approx1743$.

Before we conclude the numerical case studies, it is worth noting again that these analyses hold under the assumption that the vehicles assigned to a particular region stay there to charge. In reality, it could be the case that ride-hailing vehicles get rebalanced by the company's central operator~\cite{9838069}, so the vehicles might change the region in which they operate before charging begins. Since vehicle rebalancing also comes at a certain cost, it would be beneficial to analyze the interplay between the current model and rebalancing costs. However, this is beyond the scope of this paper and is considered an interesting direction for future research.  

\section{Conclusions}\label{sec:conclusion}
In this work, we present a novel extension of the Blotto games that, in addition to a modified Tullock contest success function, which makes the profit depend on the participation, also incorporates an additional linear cost term representing the charging costs. To analyze how a ride-hailing company wishing to enter a market should strategically distribute its fleet among different regions in the area, we start by establishing the existence and uniqueness of a Nash equilibrium for the proposed game structure. For a general case with any number of regions, we propose an analytical method to find the NE if it is located in the interior of the feasible set. Furthermore, for a two-region case, we were able to extend this method to find the NE regardless of its location within the feasible set. Finally, having established the methodology, in two numerical case studies, we illustrated the influence of charging prices on the location of the Nash equilibrium.   

n the future, we first aim to investigate if the analysis
of the Nash equilibria on the feasible set’s boundary can be extended. Additionally, we aim to broaden our investigation to encompass scenarios involving more than two players and to consider the optimization of fleet size. Finally, we plan to explore if the proposed method can be further extended to incorporate the advantages of vehicle rebalancing strategies.
\iftoggle{full_version}{\begin{figure}[!t]
    \centering
    \begin{adjustbox}{ max width=0.48\textwidth}
    \input{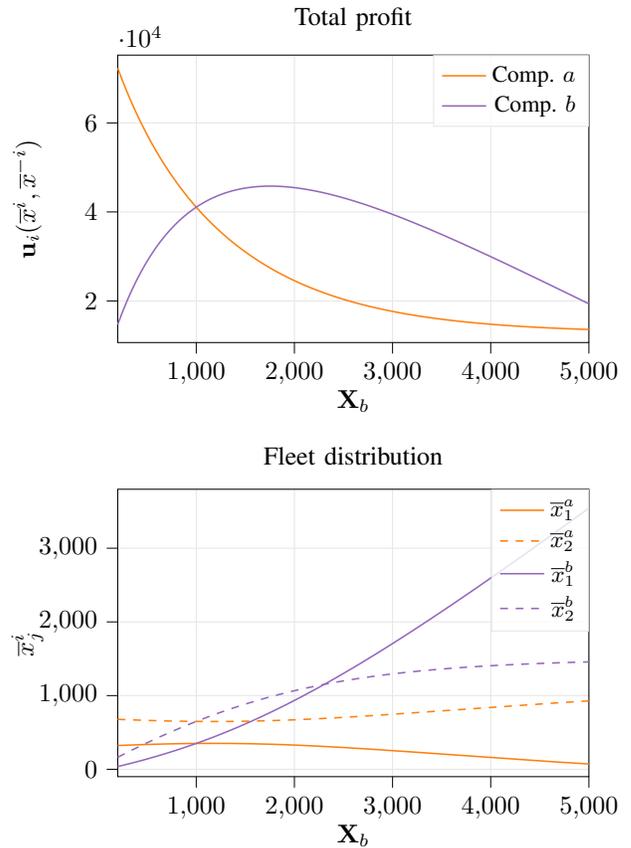}
    \end{adjustbox}
    \caption{\textbf{Two-region setup} - Attained NE of the companies and total profits of the companies for different values of $\mathbf{X}_b$}
    \label{fig:alg3}
\end{figure}}{\begin{figure}[!t]
    \centering
    \begin{adjustbox}{ max width=0.48\textwidth}
    \input{figures/FleetBchange.tikz}
    \end{adjustbox}
    \caption{\textbf{Two-region setup} - Attained NE of the companies and total profits of the companies for different values of $\mathbf{X}_b$}
    \label{fig:alg3}\vspace{-1.5em}
\end{figure}}
\bibliographystyle{IEEEtran}
\vspace{-1.5em}
\bibliography{references.bib}

\begin{thebibliography}{10}
\providecommand{\url}[1]{#1}
\csname url@samestyle\endcsname
\providecommand{\newblock}{\relax}
\providecommand{\bibinfo}[2]{#2}
\providecommand{\BIBentrySTDinterwordspacing}{\spaceskip=0pt\relax}
\providecommand{\BIBentryALTinterwordstretchfactor}{4}
\providecommand{\BIBentryALTinterwordspacing}{\spaceskip=\fontdimen2\font plus
\BIBentryALTinterwordstretchfactor\fontdimen3\font minus \fontdimen4\font\relax}
\providecommand{\BIBforeignlanguage}[2]{{%
\expandafter\ifx\csname l@#1\endcsname\relax
\typeout{** WARNING: IEEEtran.bst: No hyphenation pattern has been}%
\typeout{** loaded for the language `#1'. Using the pattern for}%
\typeout{** the default language instead.}%
\else
\language=\csname l@#1\endcsname
\fi
#2}}
\providecommand{\BIBdecl}{\relax}
\BIBdecl

\bibitem{iea}
\BIBentryALTinterwordspacing
{International Energy Agency}, ``Global {EV} outlook 2021,'' 2021. [Online]. Available: \url{https://www.iea.org/reports/global-ev-outlook-2021}
\BIBentrySTDinterwordspacing

\bibitem{9318522}
O.~N. Nezamuddin, C.~L. Nicholas, and E.~C.~d. Santos, ``The problem of electric vehicle charging: State-of-the-art and an innovative solution,'' \emph{IEEE Transactions on Intelligent Transportation Systems}, pp. 1--11, 2021.

\bibitem{EVsCharging}
Z.~Ma, D.~S. Callaway, and I.~A. Hiskens, ``Decentralized charging control of large populations of plug-in electric vehicles,'' \emph{IEEE Transactions on Control Systems Technology}, vol.~21, no.~1, pp. 67--78, 2013.

\bibitem{article4}
W.~Tushar, W.~Saad, H.~V. Poor, and D.~B. Smith, ``Economics of electric vehicle charging: A game theoretic approach,'' \emph{IEEE Transactions on Smart Grid}, vol.~3, no.~4, pp. 1767--1778, 2012.

\bibitem{9102356}
Y.~Yu, C.~Su, X.~Tang, B.~Kim, T.~Song, and Z.~Han, ``Hierarchical game for networked electric vehicle public charging under time-based billing model,'' \emph{IEEE Transactions on Intelligent Transportation Systems}, vol.~22, no.~1, pp. 518--530, 2021.

\bibitem{article5}
A.~Laha, B.~Yin, Y.~Cheng, L.~X. Cai, and Y.~Wang, ``Game theory based charging solution for networked electric vehicles: A location-aware approach,'' \emph{IEEE Transactions on Vehicular Technology}, vol.~68, no.~7, pp. 6352--6364, 2019.

\bibitem{Paccagnan2016a}
D.~Paccagnan, M.~Kamgarpour, and J.~Lygeros, ``On aggregative and mean field games with applications to electricity markets,'' in \emph{2016 European Control Conference (ECC)}, 2016, pp. 196--201.

\bibitem{Hierarchical}
M.~Maljkovic, G.~Nilsson, and N.~Geroliminis, ``Hierarchical pricing game for balancing the charging of ride-hailing electric fleets,'' \emph{IEEE Transactions on Control Systems Technology}, pp. 1--16, 2023.

\bibitem{ECC2023}
------, ``On finding the leader’s strategy in quadratic aggregative stackelberg pricing games,'' in \emph{2023 European Control Conference (ECC)}, 2023, pp. 1--6.

\bibitem{maljkovic2023learning}
------, ``Learning how to price charging in electric ride-hailing markets,'' 2023.

\bibitem{Borel1953TheTO}
{\'E}.~Borel, ``The theory of play and integral equations with skew symmetric kernels,'' \emph{Econometrica}, vol.~21, p.~97, 1953.

\bibitem{9782697}
K.~Paarporn, R.~Chandan, M.~Alizadeh, and J.~R. Marden, ``Asymmetric battlefield uncertainty in general lotto games,'' \emph{IEEE Control Systems Letters}, vol.~6, pp. 2822--2827, 2022.

\bibitem{9993253}
R.~Chandan, K.~Paarporn, M.~Alizadeh, and J.~R. Marden, ``Strategic investments in multi-stage general lotto games,'' in \emph{2022 IEEE 61st Conference on Decision and Control (CDC)}, 2022, pp. 4444--4448.

\bibitem{KIM20181}
G.~J. Kim, J.~Kim, and B.~Kim, ``A lottery blotto game with heterogeneous items of asymmetric valuations,'' \emph{Economics Letters}, vol. 173, pp. 1--5, 2018.

\bibitem{OSORIO2013164}
A.~Osorio, ``The lottery blotto game,'' \emph{Economics Letters}, vol. 120, no.~2, pp. 164--166, 2013.

\bibitem{LI2022102771}
X.~Li and J.~Zheng, ``Pure strategy {N}ash {E}quilibrium in 2-contestant generalized lottery {C}olonel {B}lotto games,'' \emph{Journal of Mathematical Economics}, vol. 103, p. 102771, 2022.

\bibitem{Rosen}
J.~B. Rosen, ``Existence and uniqueness of equilibrium points for concave {$N$}-person games,'' \emph{Econometrica}, vol.~33, no.~3, pp. 520--534, 1965.

\bibitem{Schur}
F.~Zhang, \emph{The Schur Complement and Its Applications}.\hskip 1em plus 0.5em minus 0.4em\relax {Springer New York, NY}, 2006.

\bibitem{Cauchy}
V.~Dennis~M., \emph{Cauchy's Calcul Infinitésimal}.\hskip 1em plus 0.5em minus 0.4em\relax {Springer Cham}, 2019.

\bibitem{9838069}
P.~Zhu, I.~I. Sirmatel, G.~F. Trecate, and N.~Geroliminis, ``Idle-vehicle rebalancing coverage control for ride-sourcing systems,'' in \emph{2022 European Control Conference (ECC)}, 2022, pp. 1970--1975.

\end{thebibliography}
\end{document}